\documentclass[fleqn,10pt]{wlscirep}
\usepackage[T1]{fontenc}
\usepackage{subfig}
\usepackage{lipsum}
\usepackage{graphicx,amssymb,amsmath,bm}
\usepackage{algpseudocode, algorithm}
\usepackage{enumitem,amssymb}
\usepackage{longtable}
\usepackage{array}
\usepackage{lipsum}
\usepackage{lineno}

\DeclareMathOperator*{\argmin}{arg\,min}
\newcommand{\edits}[1]{{#1}}
\newcommand{\moved}[1]{{#1}}

\title{
Building Performance Simulations Can Inform IoT Privacy Leaks in Buildings
}

\author[1]{Alan Wang}
\author[2]{Bradford Campbell}
\author[3,*]{Arsalan Heydarian}
\affil[1]{University of Virginia, Link Lab, Computer Engineering, Charlottesville, 22903, USA}
\affil[2]{University of Virginia, Link Lab, Computer Science, Charlottesville, 22903, USA}
\affil[3]{University of Virginia, Link Lab, Engineering Systems and Environment, Charlottesville, 22903, USA}
\affil[*]{ah6rx@virginia.edu}

\begin{abstract}
As IoT devices become cheaper, smaller, and more ubiquitously deployed, they can reveal more information than their intended design and threaten user privacy. Indoor Environmental Quality (IEQ) sensors previously installed for energy savings and indoor health monitoring have emerged as an avenue to infer sensitive occupant information. For example, light sensors are a known conduit for inspecting room occupancy status with motion-sensitive lights. Light signals can also infer sensitive data such as occupant identity and digital screen information. To limit sensor overreach, we explore the selection of sensor placements as a methodology. Specifically, in this proof-of-concept exploration, we demonstrate the potential of physics-based simulation models to quantify the minimal number of positions necessary to capture sensitive inferences. We show how a single well-placed sensor can be sufficient in specific building contexts to holistically capture its environmental states and how additional well-placed sensors can contribute to more granular inferences. We contribute a device-agnostic and building-adaptive workflow to respectfully capture inferable occupant activity and elaborate on the implications of incorporating building simulations into sensing schemes in the real world.
\end{abstract}

\begin{document}
\flushbottom
\maketitle
%
%

\section*{Introduction}
Due to increased awareness of energy reduction measures in buildings over the past two decades, numerous technological advancements have been introduced to monitor the changes in indoor conditions. Sensors and actuators have become increasingly integrated into buildings to reduce overall energy consumption while improving occupant comfort \cite{cureau2022bridging, wagner2018exploring}. For example, building automation systems can reduce a building's energy consumption by dimming artificial lighting when sufficient daylight is sensed in the building \cite{lu2010using}. The building can also utilize occupancy and air quality sensors to reduce energy demand by Heating Ventilation and Air Conditioning (HVAC) units in anticipation of occupants' presence or comfort \cite{yan2018evaluation}. The number of sensors installed in buildings will only grow with increasing energy prices and the known benefits of smart environments \cite{zhou2018effect}.
However, numerous challenges still exist to using sensor-collected data to improve the utility of occupants.

Firstly, and fundamentally, sensors have different data collection frequencies, so researchers cannot simply purchase any environmental sensor and install it to capture all the activity happening indoors. The frequency of data collection restricts the types of occupant behaviors that can be inferred. For instance, the Nyquist-Shannon sampling theorem demonstrates that you need to sample at more than twice the highest frequency component of the signal to correctly reverse it \cite{por2019nyquist}. The time scale differences mean that researchers cannot use a one-image per thirty-second camera to capture the behavior of subjects that occur every 15 second without losing data. Increasing the frequency, on the other end, can cause other undesirable effects, such as signal aliasing. A researcher (designer or facility manager) is still required to decide the installed sensor's specifications, such as frequency, mode, and observed variable; the very act of purchasing the hardware restricts in perpetuity the downstream observable behavior.

Secondly, sensor placements have a large effect on the downstream utility of the sensor's collected data, but positioning is often overlooked or starts off randomly then iteratively improved \cite{gautam2022efficient}. Sensors deployed at incorrect positions can result in incorrect readings \cite{ghahramani2019personal,pantelic2020personal}, but it can feel more convenient to install sensors and start collecting data as soon as possible. Furthermore, prior sensors installed for evaluating a building can become insufficient or undesireable for new uses of space. For example, some residents can move out, rendering prior sensors in locations no longer occupied redundant. Similarly, additional residents can move in, making previous coverage inadequate for the new use of space. Optimizing the position and number of sensors can result in lower energy consumption and better readings, but it is challenging to incorporate and maintain considerations of changing contexts and sensing objectives for a scaling number of sensors manually.

Lastly, in the expediency of collecting ever more data, installed sensors can inadvertently expose more information than necessary for its intended use, leading to privacy concerns for the occupants. For instance, the ``sensing by proxy" paradigm demonstrates how proxy measurements such as CO$_2$ can infer occupant count and activity\cite{jin2015sensing, szczurek2017occupancy}. Similar granular occupant activity information has been observed by other sensors as well. For example, cooking activities can be observed via the fluctuation of PM 2.5 \cite{yoon2022non}. Granular appliance use can also be effectively disaggregated via non-intrusive load monitoring \cite{huchtkoetter2020impact}. For instance, appliances such as coffee makers and hair dryers can have unique energy-use signatures relating to start-up processes and the physical makeup of the appliance. By installing a load monitor at the circuit level, the total energy use can be disaggregated from individual contributions based on the appliance's unique signatures, allowing for invasive inferences of occupant activity without needing to install any sensors inside the building itself. Other examples, such as using cameras and motion amplification, have allowed researchers to exaggerate the vibration of snack bags and reverse engineer decipherable sounds using visual data \cite{wadhwa2016eulerian}. 

As more projects utilize machine learning and other computational methods to retrieve sensitive data from the indoor environment, we instead consider if similar computational methods exist to help reduce inferable information from sensors and protect the privacy of building occupants. One promising avenue that can help sensor installations navigate the potential overreach of sensing is at the intersection of simulations and sensor placements \cite{wagiman2020new}. Simulations have traditionally been used to assess different building performance attributes during the design phase. The orientation of buildings and placement of windows, for example, can be explored and quantified using a score called daylight autonomy \cite{lm2013approved}. Simulations of the weather conditions, and movement of the sun, in conjunction with the location of the building, size, material, and orientation of the window and room, enable architects to uncover the total amount of time over a whole year when daylight can effectively stand in for artificial lighting. Similar simulations and metrics can be found for HVAC, where given the hours that the building will be occupied, the room size, the expected occupancy, the total energy use, and required airflow can be predicted and quantified \cite{shin2019thermal}. The acoustic qualities of a building can also be designed and tailored to better match the intended use of space, such as longer reverberations for music halls and shorter reverberations for classrooms \cite{wang2022development}. \edits{The timing in the pipeline during which these simulations are used represents a fundamental discourse in digital twins \cite{batty2018digital}. Specifically, a simulation model can help optimize the placement of sensors to inform on occupant activity,  and it does not need to be run in real-time parallel to the physical twin for it to have a lasting impact on the smart environment.}

Compared to manual or autonomous methods of sensor position optimization, where robots are used to sample the environment routinely and iteratively uncover the most optimal positions for the static environmental sensors \cite{belhaj2016smart}, simulations enable a low-cost alternative for testing unlimited virtual sensors positions at the cost of computational power. Furthermore, instead of navigating protocols for institutional review boards (IRB) or logistics of the environment (e.g., to avoid a party or speaker event in the building), simulations enable researchers to avoid a broad range of complexity that can cause nontrivial disruptions for both study administrators and study participants. Researchers (designers or other decision makers) can run unlimited what-if scenarios to see how different environmental or user-related factors may impact the changes in the downstream signal before interacting with the physical environment. For example, the movement of occupants can be simulated using artificial agents to assess the ease of navigating the space in case of emergencies \cite{li2014bim}, the sun's movement can be simulated using historical sky information \cite{aguilar2022validation}, and stochastic models can be used to simulate an occupants interaction with building controls \cite{reinhart2004lightswitch}.

In this paper, we demonstrate the potential of physics-based simulation models to quantify the minimal number of positions necessary to capture sensitive inferences. We show how a single well-placed sensor can be sufficient in specific building contexts to holistically capture its environmental states and how additional well-placed sensors can contribute to more granular inferences. Specifically, we focus on lighting simulations as a test case because of its accessibility and geometrically-dependant nature. We answer two research questions: \begin{itemize}
    \item \textbf{RQ1}: What are the challenges in translating lighting inferences found in the real world to inferences found in the simulations? And,
    \item \textbf{RQ2}: How can lighting simulation be used to inform us about how informative a set of light sensors positions are given some assumptions about space-use?
\end{itemize} 

We advocate for the use of simulation as a standard tool for 1) identifying the ideal location for sensors and minimizing the number of sensors distributed and 2) identifying potential information that sensors can collect when deployed in real life by showing the capabilities of simulations to calculate building states containing granular occupant activity exhaustively. In other words, we show how the simulated environment allows researchers to assess the effects that the position of sensors and the geometry of the building can have on occupant activity inferences. The workflow demonstrates an avenue for future researchers to verify the possible inferences of existing sensor positions or use the ``adjustment of sensor positions" as a method to limit sensor inference overreach.

\moved{
\section*{Methods} \label{sec:methods}
We consider the scenario where a researcher is trying to ascertain the light state of on and off for individual luminaires. Given the additive nature of light, we approach decomposing the summed light contribution at a sensor point in the building by formulating it as a \textit{Perfect Sum Problem} with a noise threshold $\epsilon$. Since light intensity diminishes equal to the inverse of the square distance from the source, and each luminaire has photometric data, moving a light sensor's final placement at a variable distance away from light sources can enable individual contributions to be disaggregated. For instance, we can utilize the geometry and drop-off of light intensity by distance to coordinate unique fingerprints for each luminaire in range. Furthermore, having the ability to utilize the entire building space as potential placement areas enable researchers to exhaustively explore the ability to utilize \textit{sets} of potential sensor positions that traditionally require repeated trial-and-error to capture. 

\paragraph{State Inferences}
Let $L= [l_0, l_1, ... l_{n-1}]$, where $L$ is a configuration vector of $n$ individual light source states $l_i\in\{0,1\}$. Then, for any location $s$, given a configuration $L$, we see the contribution vector $X(s,L) = [x_0, x_1, ..., x_{n-1}]$, where the contributions from each light source $x_i$ corresponds to each light state $l_i$ modified by distance and obstruction. The maximum number of possible configurations is then equal to the cardinality of the power set with each light source, or $2^{n}$.


Given a perfect sum solver, we can take a target sum $\text{\textbf{K}}$, threshold $\epsilon$, and list of individual contributions $[x_0,x_1,...x_{n-1}]$ and return a list of lists that contain all possible combination of contributions that add up to $\text{\textbf{K}}\pm \epsilon$. Each list corresponds to a possible configuration that fits the constraints, but since ultimately only one configuration is correct, we calculate the accuracy for each inference using
$
  \textnormal{Accuracy} = \frac{|L \cap L_{\text{infer}}|}{|L \cup L_{\text{infer}}|}
$
and return the mean. 

Sensor readings from multiple points do not always corroborate to the same lighting configurations. To overcome this, we disambiguate misaligned light configuration inferences by using a voting vector $ V(s,L) = \begin{bmatrix}
    v_{0}, v_{1}, ... v_{n}
\end{bmatrix}$, from each sensor, where $v_{i}$ is 1 if the luminaire is determined to be on, -1  if the light is determined to be off, and 0 if the light cannot be detected (i.e., the sensor is out of range of the luminaire). The final inference is chosen based on the summed votes. If the value is greater than zero, the luminaire is on. If the final value is less than zero, we consider the luminaire off.

\paragraph{Distinctness Score}
To generalize the real-world sensing to building simulations, we introduce an error threshold $\tau$, and define a distinctness vector $D_{\tau}$ as:

 \begin{equation}\label{eq:d}
    D_{\tau}(X(s,L)) = 
        e_{0}, e_{1}, ..., e_{n-1}, \text{where } 
     e_{i} = \begin{cases}
    1, & \text{if} \; \argmin_{j} |x_i - x_j| > \tau, i\neq j\\
     0, & \text{otherwise}.
   \end{cases}
\end{equation} 

For example, given a contribution, $X(s,L) = [1, 2, 4]$, if the error threshold $\tau = 1$, then $D_{1}(X(s,L)) = [0, 0, 1]$ and the distinctness score  $\text{\textbf{D}}_{1} = \sum{D_{1}} = 1$. However, given a $\tau = 0.5$ for the same contribution, $D_{0.5}(X(s,L)) = [1, 1, 1]$, so $\text{\textbf{D}}_{0.5} =3$. The sum of the distinctness vector helps us decipher the total number of detectable lighting states at a given position and accounts for the sensor's resolution when assigning credit. We utilize this score in our simulations, where the virtual light sensors in simulation do not have the added lumen degradation and other measurement noise terms.

For our activity inferences, we account for $j=3$ door angles (i.e., $\{0^{\circ}, 45^{\circ}$, $90^{\circ}\}$), for $m=2$ doors. To find the distinctness score for each given application state $a$ is then: 
\begin{equation}
    \text{\textbf{D}}  = \sum_{p=0}^{2^n-1}\sum_{q=0}^{jm-1}{D(X(s, L_p, a_q))}
\end{equation}

Because choosing the minimum combination of sensor locations that can detect all possible applications is a known NP-Hard problem called the Minimum Set Cover (MSC) Problem, for simplicity, we visualize the inferable states for only the single-sensor scenario. Finally, we show an example set of light sensor locations that can capture our latent variable and the building light states using a known algorithm for the MSC problem: the Greedy Set Cover Approximation (GSCA) \cite{young2008greedy}. To see if a set of light sensors capture the dynamic building information, we can then use the entire application states as the \textit{universe} of states to cover a threshold $<\tau$ against all other sensor locations rows for the same application state column to inform \textit{membership}, to find the minimum subset of sensor locations required to cover the entirety of the application states. \edits{We target door usage because of its fixed nature. The opening and closing of doors have a large effect on perceived lighting by the sensors, with the potential to block off sections of lighting altogether consistently. Further, door usage is a deeper insight into space usage not expected by light sensors. Using the bathroom, bedroom, or living room naturally involves opening and closing lights and doors. In this case, where the lighting does not automatically turn off, being able to discern door movement even while all lights are on allows more granular inferences about space use to be made.}

\subsection*{Experiments}
As shown in Figure \ref{fig:door_test}, even for a single light source, the angle of a door's rest state can significantly impact the final detected light signal. Using these methods, we conducted two experiments: 1) we deployed light sensors into the modeled residential setting to explore real-world challenges with detecting light states and fusing inferences from multiple sensors together, and 2) we modeled the residential building and simulated a set of dynamic building elements to analyze the number of light states and latent states that can be observed. The real-world study was conducted first as a sanity check; if we can arbitrarily place sensors using only human intelligence, it is unnecessary to utilize simulations to support the sensor placements. The simulation study was conducted to see whether the simulation space would corroborate with the findings found from our real-world study and also to gauge the minimum number of sensors that can capture a set of luminously disruptive behavior.

\begin{figure}[!ht]
    \centering    \includegraphics[width=0.7\linewidth]{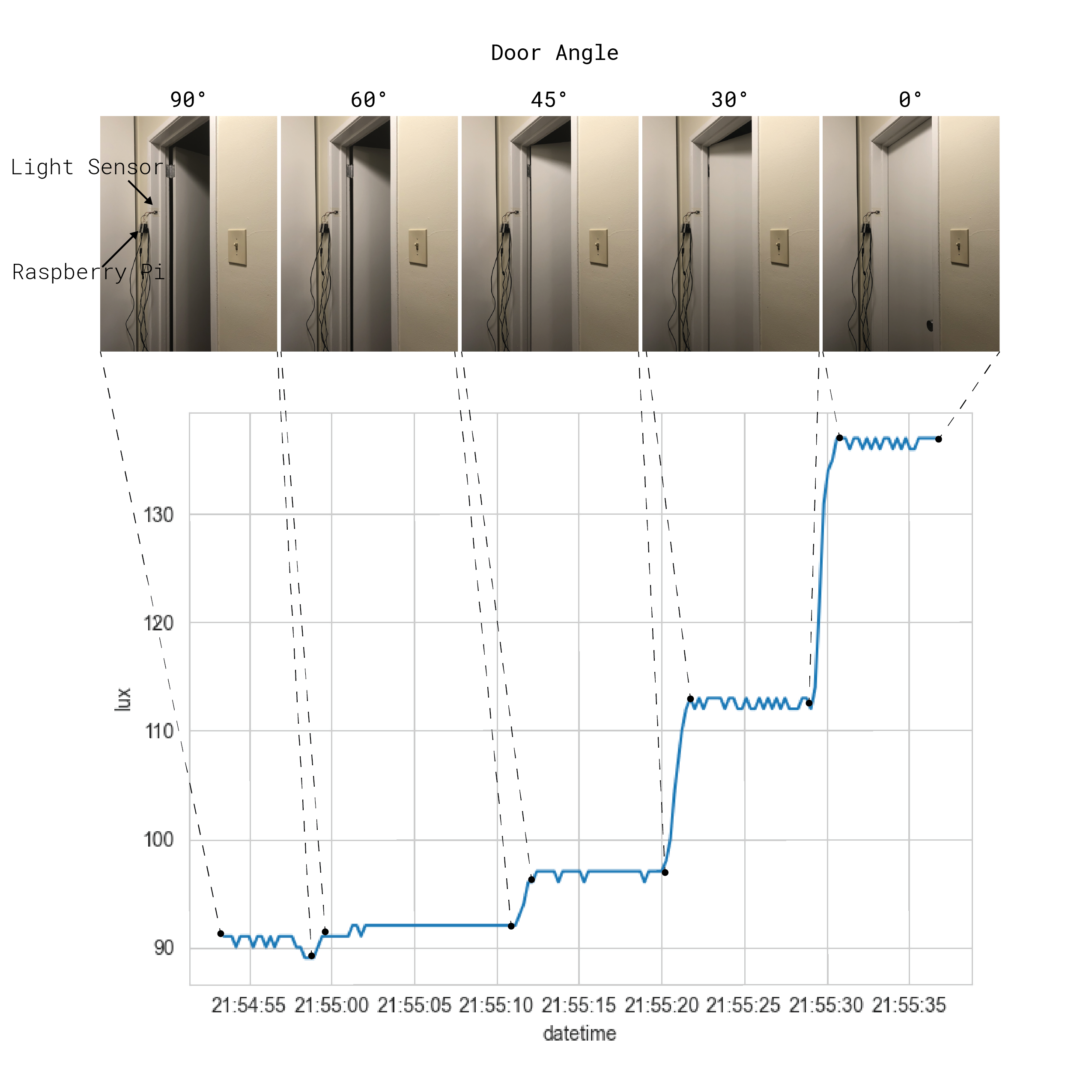}
    \caption{Relationships between building elements and lighting can be used to inform about changes in the physical environment.}
    \label{fig:door_test}
\end{figure}

\paragraph{Real-world Experiment}
In the real world, we retrofitted all lamps in a residential apartment building with 800 lumens 10 watts Philips Hue A19 Lamps and grouped each set of lights to their corresponding luminaire. For example, two to three lamps can be associated with each light switch. We controlled each Philips Hue lamp using the Hue API and Python to limit the need to alter the light switches manually. We then used Raspberry Pi 4s connected with CQRobot TSL2591X light sensors with an effective sensing range of 0 to 88,000 Lux as our sensor, communicating using the I2C interface with the Pi. To account for the jitters, we used the mean lux values for three seconds as the baselines, collected at 4.7 Hz after our code changed the lighting configurations configuration $L$ for three seconds because of the changes in light intensity during state transitions. We used the proximity of the light sources as a guide to installing the light sensors in each location and orientation, as shown in Figure \ref{fig:lighting-plan}. Specifically, we visually looked for sensor positions on different walls, enabling the sensors to capture light from different light sources. Then, we moved our sensor system across each position and automated the lighting transitions using the Raspberry Pi, exporting the final data into a CSV file for post-processing. The final accuracy we report is explained in the results, where the ground truth is the input command we used to automate the light states. \moved{To install an initial set of sensors, we identified walls in the testbed that all the luminaries can reach and then placed seven sensors on those walls, seven feet above the floor shown in Figure \ref{fig:lighting-plan}. The idea is to see if it is possible to install sensors in locations that avoid the noise in the data caused by shadows in human traffic, the reflection of furniture, and other LEDs from appliances and objects. This allows us to accurately detect the lighting state of the building for the static open door scenario before we dive into permutations of doors in the simulation experiment.}

\begin{figure}[!htp]
    \centering
    \includegraphics[width=0.5\linewidth]{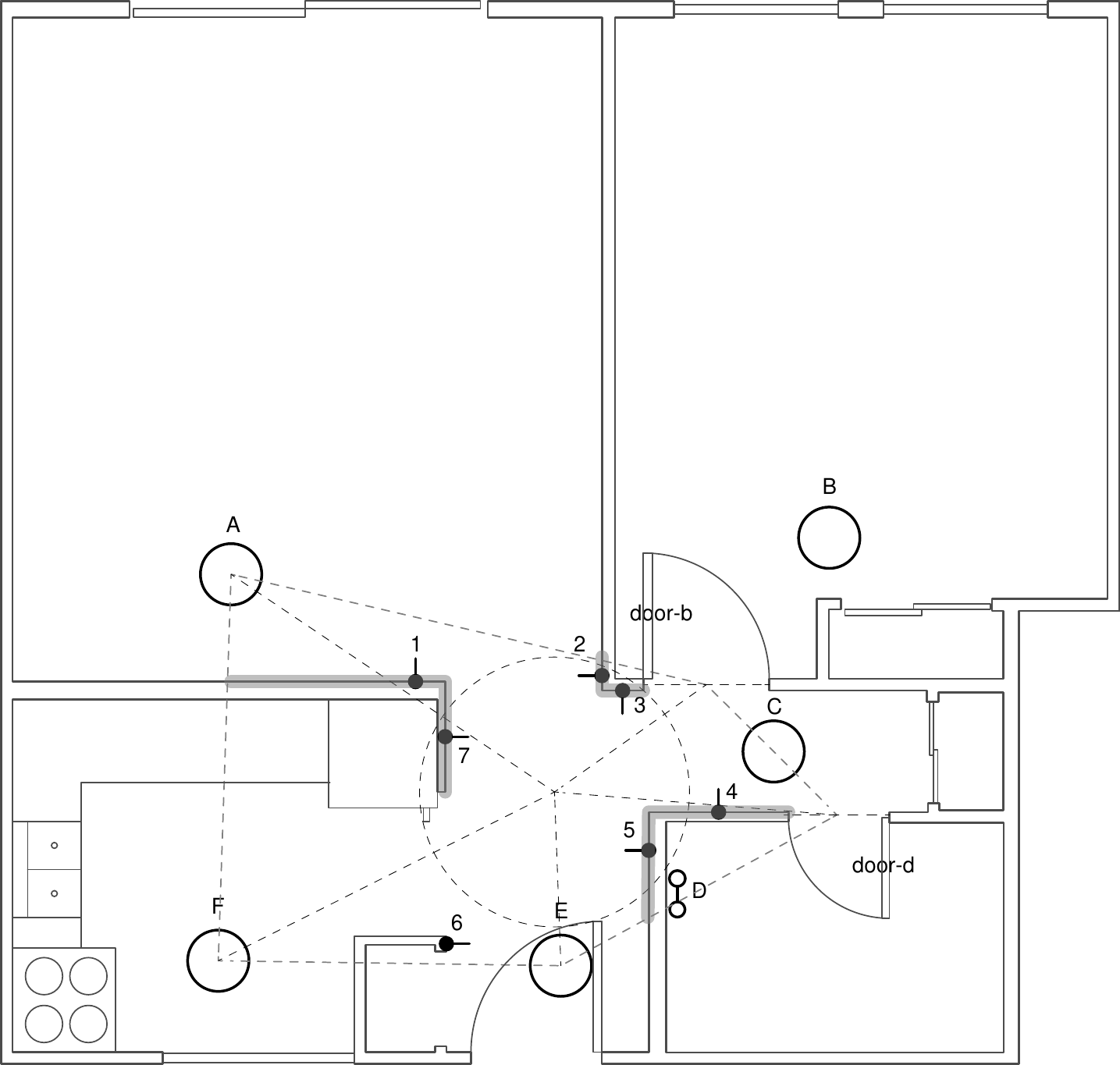}
    \caption{Building and Lighting Layout. The numbers denote the light sensor positions, and the alphabets denote the light sources. The dashed lines denote geometric boundaries we used to search for our candidate walls to install the sensors.}
    \label{fig:lighting-plan}
\end{figure}

\paragraph{Simulation Experiment}
In a simulation, we utilized grasshopper to parameterize multiple door movements and simulated each of the allotted door angle combinations using Rhino \cite{cheng2013inside}. We then used the grasshopper plug-in honeybee \cite{roudsari2013ladybug} as we have in our previous work \cite{heydarian2017towards} to extract the lighting contributions at each sensor point. In essence, the plug-in acts as a middleware that takes the building geometry and photometric lighting files (IES files) that describe the geometric intensity distribution of light and passes them into the lighting render engine radiance \cite{ward1994radiance}. \edits{The experiment was assumed to be undertaken at night with covered windows to avoid external light sources. We used the simulation default material for all wall, floor, ceiling, door, and window objects. For the photometric lighting files, we used generic flush-mounted dome lights for fixtures A, B, C, E, and F and a generic wall-scone fixture for D.} We automate the inputs to the grasshopper workflow and export text files we convert for post-processing using Python modules inside grasshopper. The process returns a ray-traced light rendering of the building for each virtual sensor position defined. We extract each lighting value for post-processing by passing it through the Perfect Sum solvers and the GSCA algorithm. This workflow enabled us to explore potential lighting differences in the overall environment without needing to physically place new sensors, change which luminaires were on and off, and move the doors in the real world.
}

\section*{Results}\label{sec:results}

We first conducted a feasibility study to explore the potential challenges of collecting sensor data \edits{manually}. We then used the experiences we have learned to inform our modeling of the building in simulation space. We summarize the real-world experiment and the simulation experiment below.

\begin{figure}[htbp!]
    \centering  
    \includegraphics[width=0.5\linewidth]{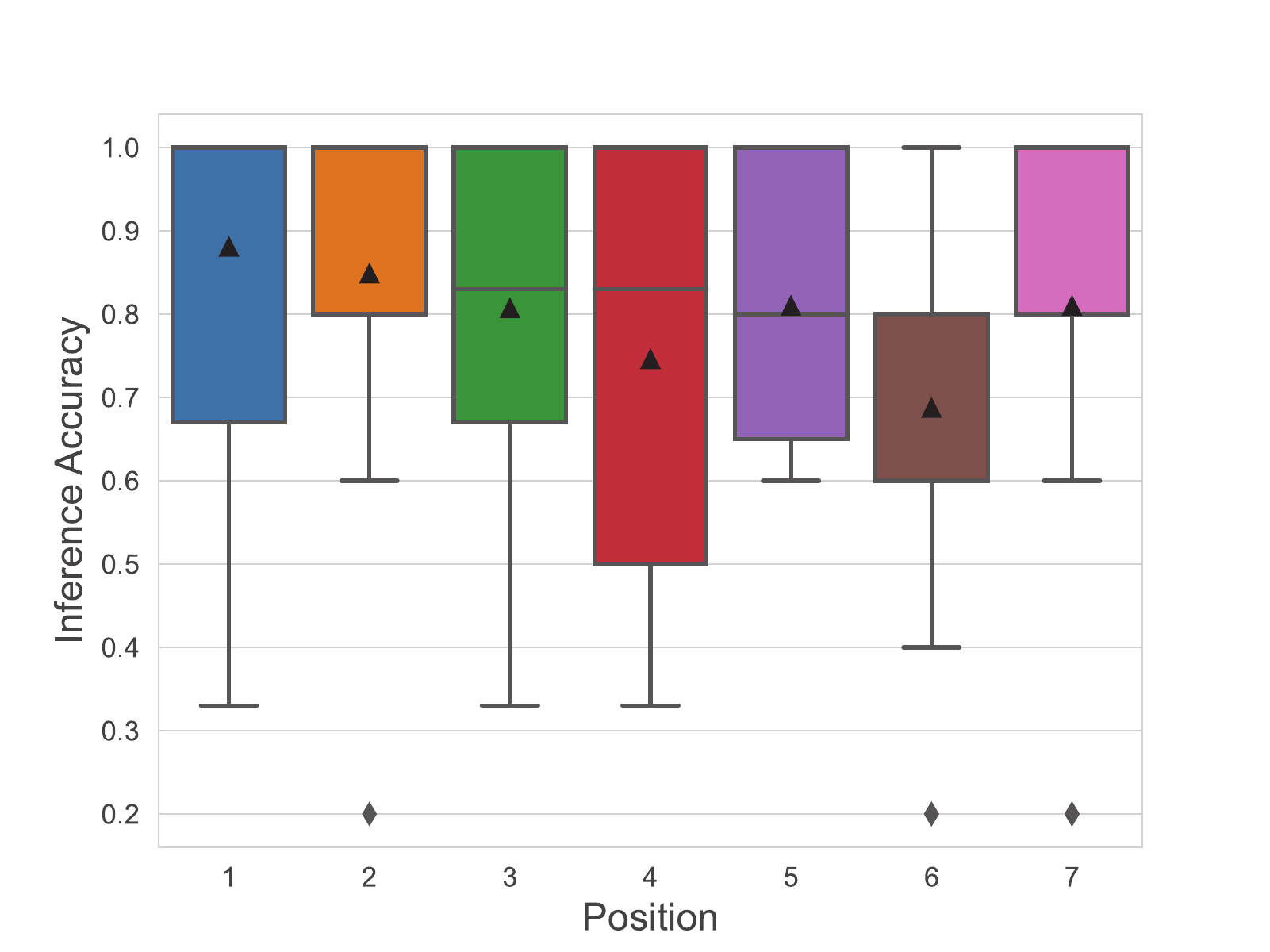}
    \caption{Light state inference accuracy for each of the 7 positions for 64 possible lighting states for manually selected positions. The triangles mark the mean accuracy for each sensor location, \edits{while the line across the middle of the box marks the median accuracy.}}
    \label{fig:sensor_accuracy}
\end{figure}

\paragraph{Real-world Experiment}
Figure \ref{fig:sensor_accuracy} shows the spread of individual accuracy for those locations. The median accuracy of inferences was generally above 80\%, with location 6 having the lowest median accuracy. We suspect this due to the direction the sensor is facing being directly opposite to luminaire F and reflected lights off the wall being less potent than the drop off of intensity due to distance.  Overall, we still found it possible to disaggregate all possible light states using a single sensor (i.e., a light sensor located at position 4), but many factors can contribute to the imperfect inferences. For example, when an occupant closes the door to any room, the information about the light state in that room is lost to sensors outside. We also experience situations where the inferred light state of the signals does not align with each other, because the noise in the environment and sensor was larger than the resolution required to differentiate the states. For example, the lighting contribution from two different light sources can add up to be the same (e.g., 2+5 and 3+4 both equal 7) leading to ambiguous readings. This led us to utilize a voting mechanism to reduce the overall error of the system, which is further described in the methods section.

When we automated the lighting states, we also realized that, unlike what may happen in simulations, the switching on and off of lights in the real world is not instantaneous. Specifically, when turning on the lights, there is a distinguishable start-up time when the light starts dimmer after the switch is flipped and approaches its final brightness after a delay. Furthermore, when retrofitting the luminaires, we noticed that not all of the lights were using the same bulbs and that likely more frequented areas had bulbs that were more frequently replaced. This suggests that keeping track of the light usage in a building can also be helpful to track which lights might need to be replaced and account for the lamps' brightness degradation over time. Finally, we found that the number of available sensors, microcontrollers, and outlets also impose a limit to how many positions can be tested at once. In addition to purchasing a long extension cord to move our sensing apparatus across the building, we also had adhered to a procedure of moving sensors, installing sensors, running through all the light states, and uninstalling sensors to sense each position.

\begin{figure}[!htp]
    \begin{tabular}{cccc}
    \subfloat[Bed Door $90^{\circ}$, Bath door $90^{\circ}$ (open-door scenario)]{\includegraphics[width = 0.22\linewidth]{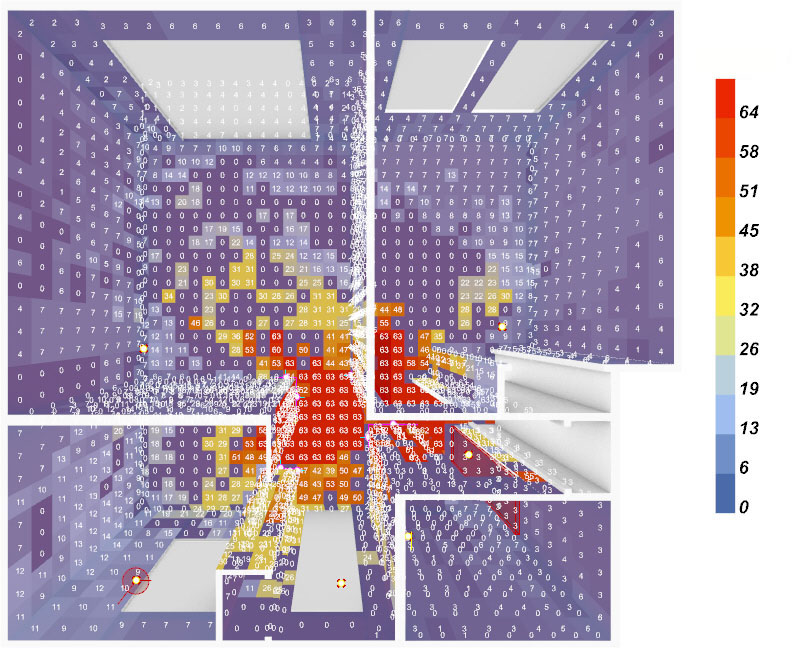}\label{subfig:a}} &
    \subfloat[Bed Door $45^{\circ}$, Bath door $90^{\circ}$]{\includegraphics[width = 0.22\linewidth]{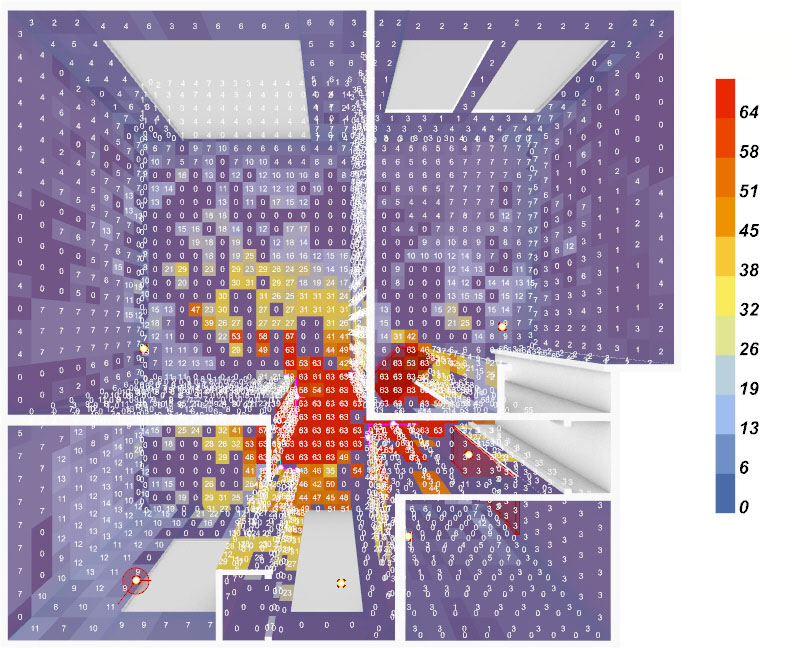}\label{subfig:b}} &
    \subfloat[Bed Door $0^{\circ}$, Bath door $90^{\circ}$]{\includegraphics[width = 0.22\linewidth]{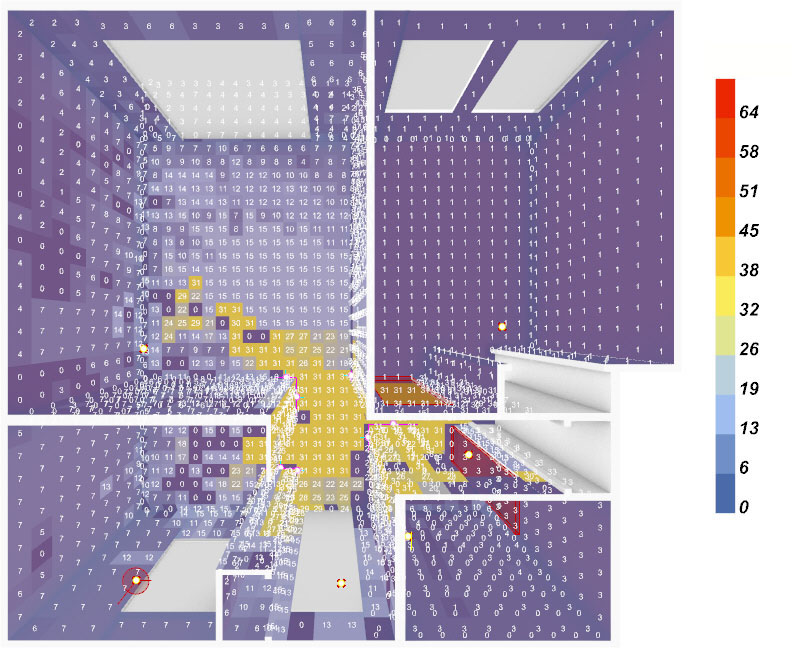}\label{subfig:c}} &
    \subfloat[Bed Door $90^{\circ}$, Bath door $45^{\circ}$]{\includegraphics[width = 0.22\linewidth]{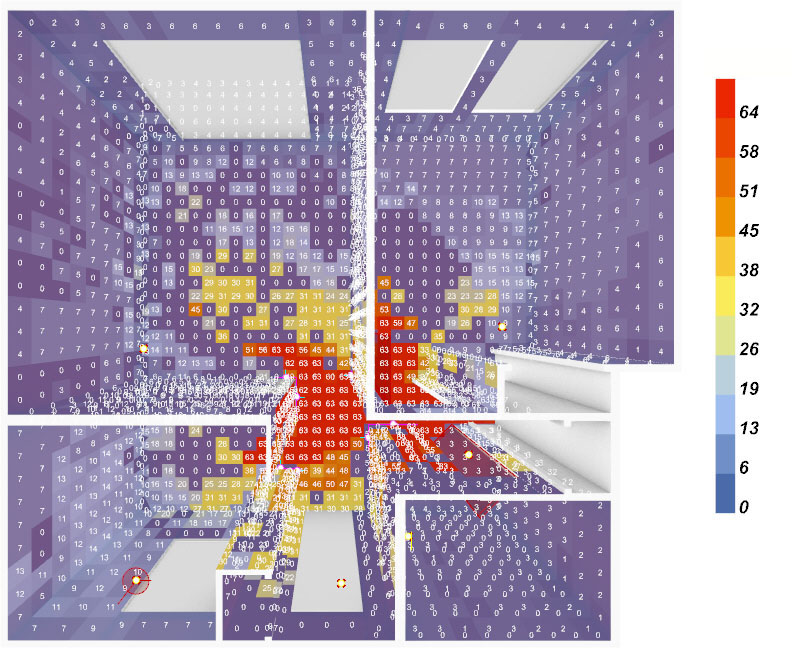}\label{subfig:d}} \\
    \subfloat[Bed Door $45^{\circ}$, Bath door $45^{\circ}$]{\includegraphics[width = 0.22\linewidth]{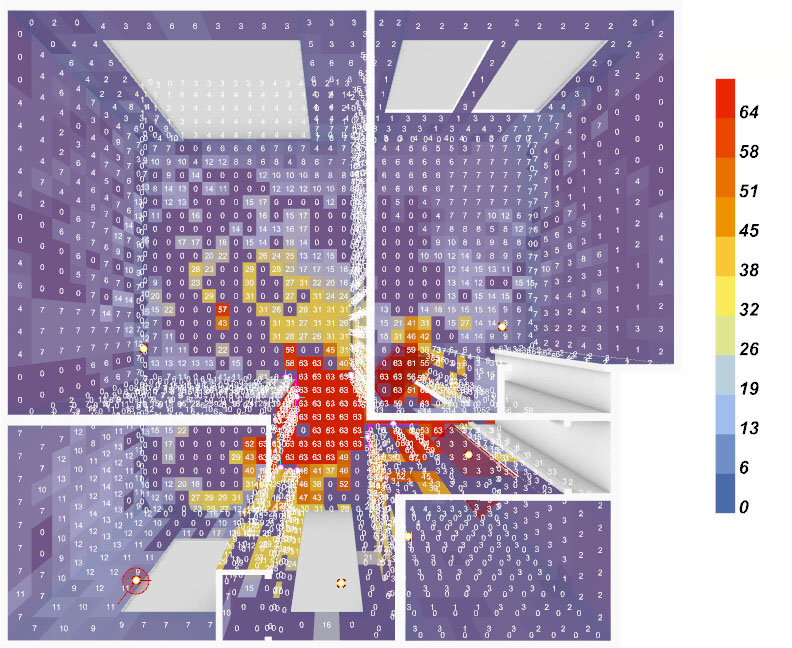}\label{subfig:e}} &
    \subfloat[Bed Door $0^{\circ}$, Bath door $45^{\circ}$]{\includegraphics[width = 0.22\linewidth]{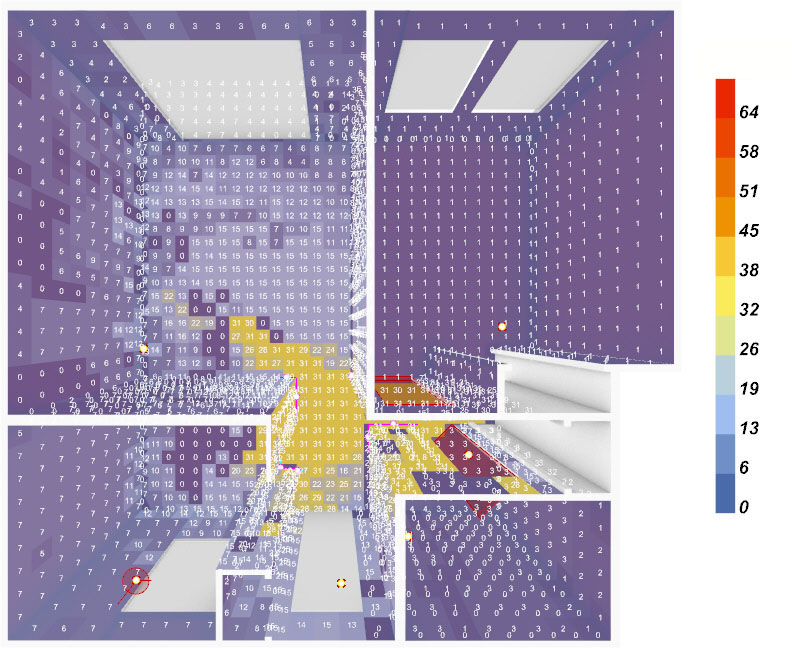}\label{subfig:f}} &
    \subfloat[Bed Door $90^{\circ}$, Bath door $0^{\circ}$]{\includegraphics[width = 0.22\linewidth]{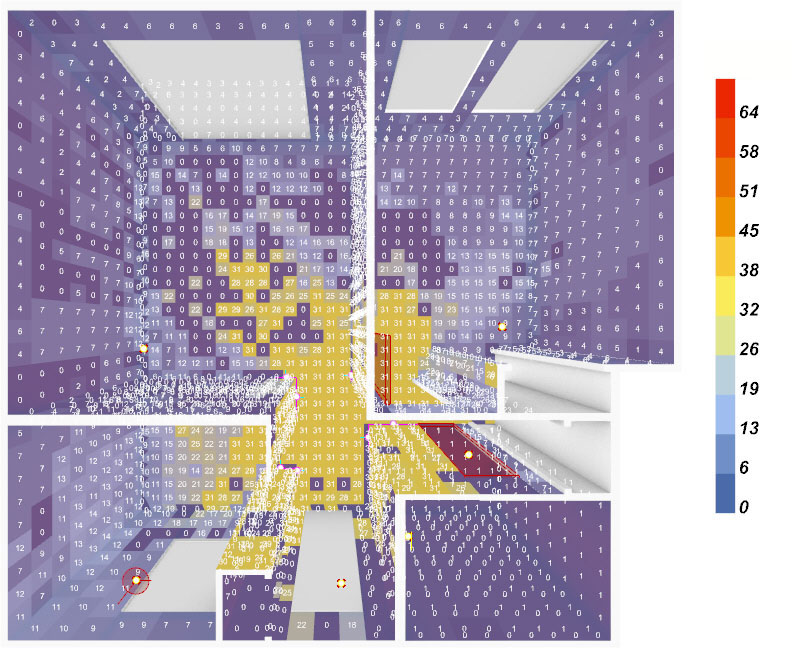}\label{subfig:g}} &
    \subfloat[Bed Door $45^{\circ}$, Bath door $0^{\circ}$]{\includegraphics[width = 0.22\linewidth]{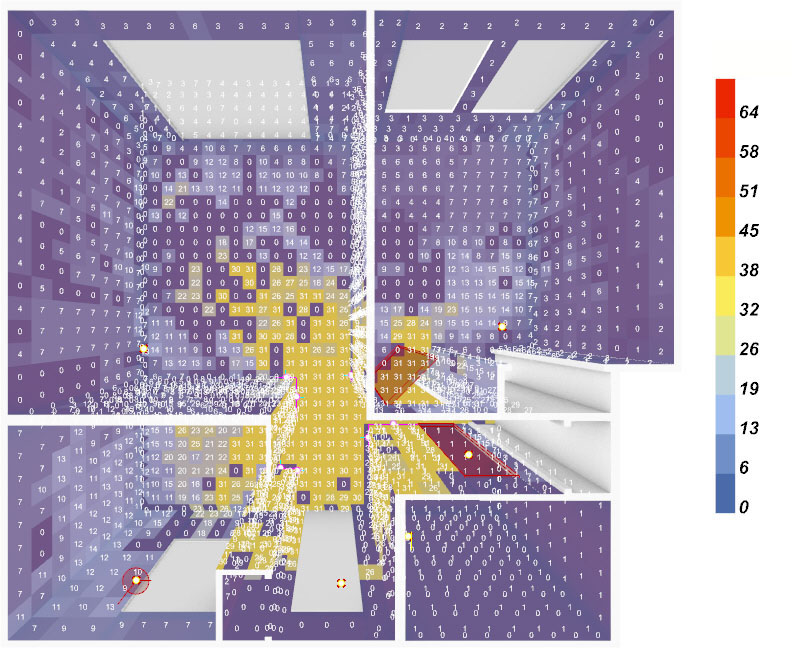}\label{subfig:h}} \\
    \subfloat[Bed Door $0^{\circ}$, Bath door $0^{\circ}$]{\includegraphics[width = 0.22\linewidth]{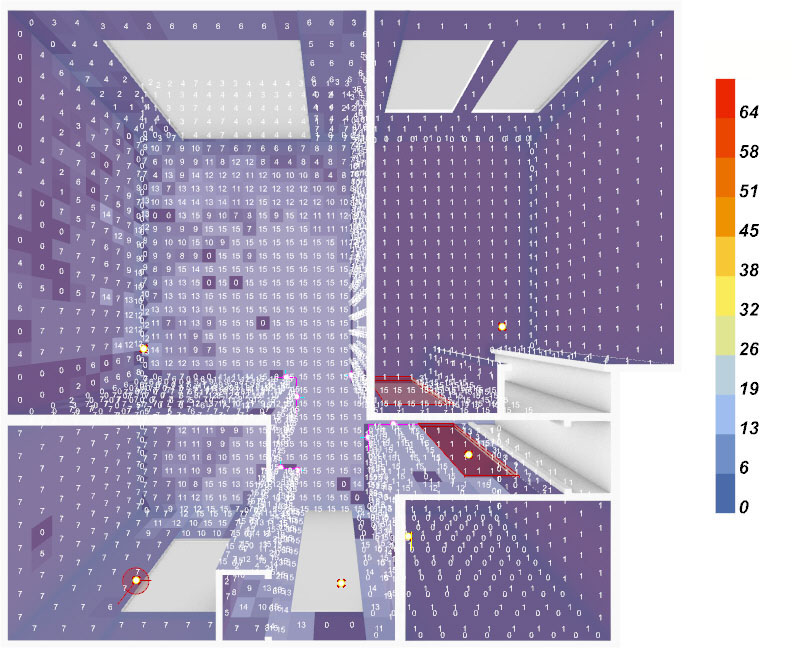}\label{subfig:i}} &
    \subfloat[Aggregate latent variable privacy $\text{D}_{0.01}$]{\includegraphics[width = 0.22\linewidth]{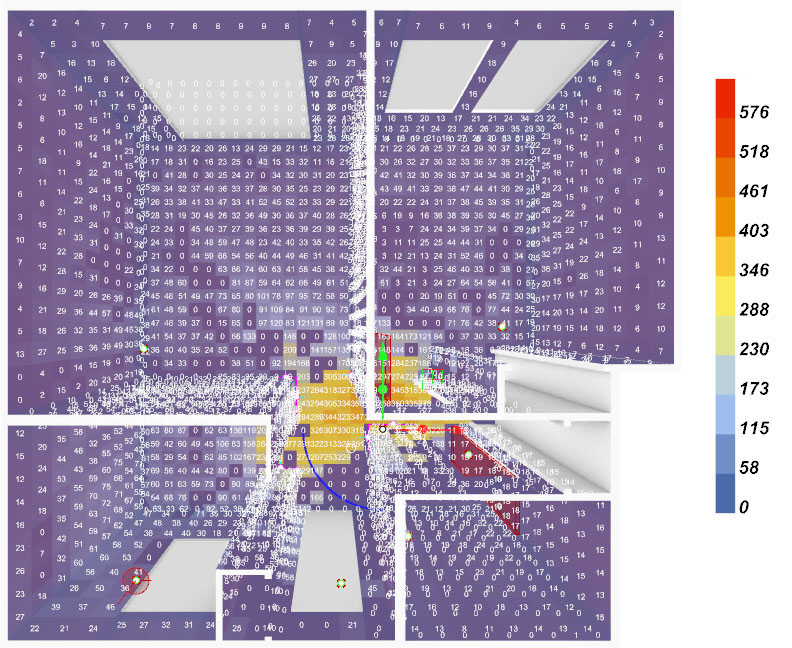}\label{subfig:total}} &
    \subfloat[One light sensor can sense all 64 light states for the open-door scenario using GSCA]{\includegraphics[width = 0.22\linewidth]{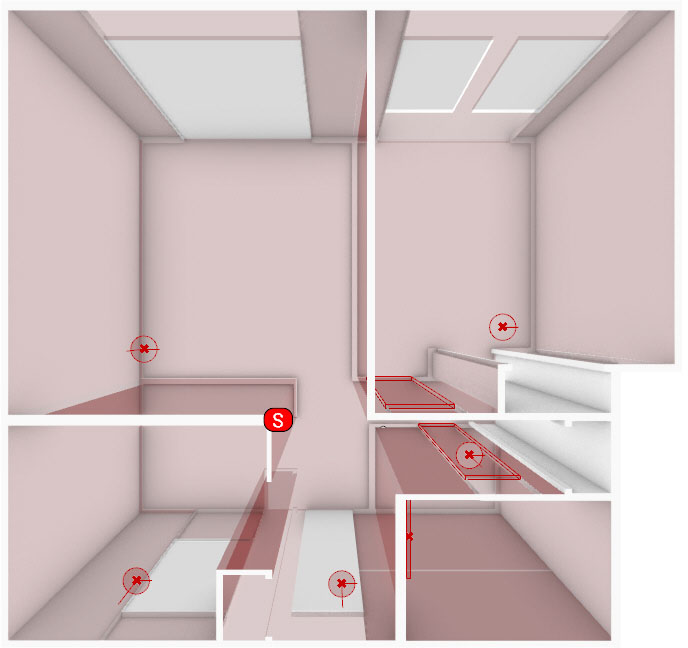} \label{subfig:GSCA-light}}  &
    \subfloat[31 sensors are needed to sense a total of 576 states using GSCA for the dynamic-door scenario]{\includegraphics[width = 0.22\linewidth]{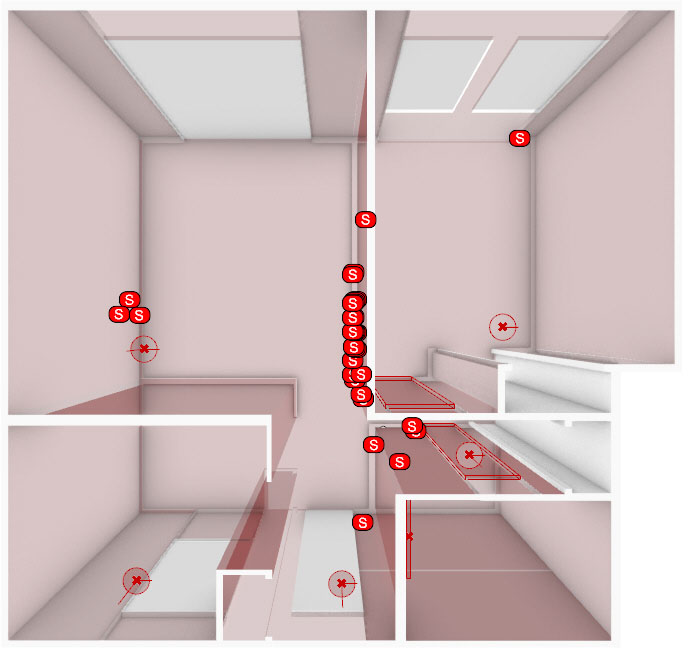} \label{subfig:GSCA-door}} \\
    \end{tabular}
    \caption{Compilation of the distinctness score $\text{\textbf{D}}_{0.01}$ heat maps for each building configuration. In addition to the resolution of the sensor, the building's physical configuration can also systemically alter the inferable states of the sensor set. \edits{Note that the scale for Figures \ref{subfig:a} through \ref{subfig:i} is on a scale out of 64, while the aggregated scale on Figure \ref{subfig:total} is out of 576.}}
    \label{fig:sims_tau0.01}
\end{figure}

\paragraph{Lighting Simulation Results}
Utilizing the information we've learned from the real-world experiment, we developed a lighting simulation that deviated from traditional lighting simulations to explore locations where static sensor installations can give us the most information. Specifically, in our lighting simulation, we included \textit{walls} instead of using the traditional work plane--an imaginary plane set at the level of a desk where work is done--because we are not interested in the utilization of the space but rather in the ability to install real static sensors and detect different behaviors in the space (e.g., light switch behavior). The results from the simulation are shown in Figure \ref{fig:sims_tau0.01}. Figures \ref{subfig:a} through \ref{subfig:i} represent the different states of dynamic building elements, with the heatmap showing the distinctness score $\text{\textbf{D}}_{0.01}$ value from 0 to 64, representing the number of collective states a sensor placed at the location can detect. Figure \ref{subfig:a} represents a typical light sensing simulation, where the movement of additional building elements is not considered. In this scenario, many positions in the middle of the room where all the lights can reach can be used to infer the light state of all luminaires. From Figure \ref{subfig:b} to Figure \ref{subfig:i}, we show that these informative middle positions diminish as the doors close and block out lighting contributions from different sources. In Figures \ref{subfig:d} and \ref{subfig:e}, when all doors are still partially open, we see a slight reduction in areas that can still make all targeted inferences. However, in Figures \ref{subfig:c}, \ref{subfig:f}, \ref{subfig:g}, \ref{subfig:h}, and \ref{subfig:i}, we see the total possible inference visibly diminish by half to three quarters.

Figure \ref{subfig:total} represents the most informative single location accounting for all possible door states. As much as there are informative locations in the middle, we also observe there to be spots of lost information, dark zones that have lower inference potential. This is a result of the clashing of light contribution combinations that lead to ambiguous readings. Comparably, while Figure \ref{subfig:i} has lower number of possible inferences in the middle, there are also less informative ``dark spots" as a result of collisions. Figure \ref{subfig:GSCA-light} shows how Greedy Set Cover Approximation (GSCA) found a single location that can detect all light states, but Figure \ref{subfig:GSCA-door} shows that the previous best location is no longer valid when we account for the opening and closing of doors. More specifically, we can see a set of sensors being placed deep into rooms away from noisy areas in the center, which help to disambiguate the readings when the lighting signals are muddied by reflection and attenuation in the center area. 

\section*{Discussion}
\paragraph{Key Takeaways}
The key takeaway from our study is that: \textit{sensor position is important, and simulations can be used to quantify just how important position is}. \edits{By quantifying inferable information in simulation, building operators can adjust the privacy-utility spectrum for where sensor installation should occur before deployment. The quantity and location of sensors can be altered to purposefully remove possible inferences based on the physical attributes of the environment. Even after deployment, simulation elucidates where the current installation is on this spectrum and ways to navigate this trade-off. In our scenario, this trade-off between privacy and utility is exactly the distinctness score, a quantifiable value between zero and the total number of states we are considering.} Towards answering \textbf{RQ1}: we found that the informativeness of the sensor location also hinges on the sensor's resolution. Paradoxically, the more contributions from different light sources sensed by a sensor-location pair, the more chance there can be ambiguous readings in the perfect sum solver because of the number of possible combinations and jitters in the sensor signals. \edits{These jitters, shown in Figure \ref{fig:door_test}, signify a vital distinction between real-world and virtual sensors in simulations. Adding more sensors in the physical world does not, by default, increase the accuracy of the final light state inference. The requirement for accurate sensing relies on the majority of the votes cast being accurate. To know the lowest level of resolution permissible for accurate sensing before purchasing a sensor, simulations can be a useful tool to assist with planning. In the simulation, the resolution of the virtual sensors is deterministic under the same parameters, and adjustments can be made to simulate different sensor resolutions by introducing additional noise terms. Further, by adding more information about the activity, simulations can be improved to account for different sampling frequencies by discretizing sample points based on a continuous response function. With sufficient computational resources, permutations of different sampling policies and sensor descriptions can be used to optimize the inference accuracy, redundancy, and efficiency. With simulations, researchers can achieve more intrusive inferences with fewer sensors, fewer samples, and less energy compared to without simulations.}

Towards answering \textbf{RQ2}: we found that utilizing lighting simulations with a formulation of \textit{Perfect Sum Problem} with the \textit{Set Cover Problem} allowed us to quantify the minimal number of light sensors that are required to capture the light state of the building, including modification of the doors. We found that as long as the sensor is placed in a location where all light sources can reach and result in a different contribution, a single sensor is theoretically enough to infer all of the possible light configurations in the building if the resolution equivalent $\tau$ is sufficiently small. However, as doors are introduced that can block off lighting contributions from other luminaires, the minimum number of sensors required to sense the lighting state of the building becomes equal to the total number of independent zones. For example, three separate rooms require at least three sensors to detect the lighting state, regardless of the number of luminaires in each room. The minimum number of sensors required increases from one to 31 when we accounted for the movement of the doors as seen by Figure \ref{subfig:GSCA-light} compared to Figure \ref{subfig:GSCA-door}. This indicates that even simple residential buildings with no dimmers can result in complex luminous environments if commonplace building elements such as doors are considered. Unless researchers had thousands of sensors, placed at every inch of the space, they wouldn't be able to test the entirety of the space at once.

Finally, simulations can be much faster at the search for optimal positions. In our experiment, it took us roughly 30 minutes to test one position in the real-world, where as in simulation space, we cover roughly one sensor location every 0.11 minutes of simulation (about 2,800 points can be calculated every 5 minutes for the 64 different light states). This approximates a 270-times increase in efficiency when using simulations to test positions compared to real-world testing, not accounting for set-up time in either scenario. Because simulations also do not require the researchers to be physically present in space, simulations hold a distinct advantage over manual testing as an important step to enhance physical sensor deployments. With increased simulation scenarios that incorporate human movement and other modalities such as noise and HVAC, the digital space will increasingly become more critical not just for sensor positioning but also for a large myriad of selection tasks. From what type of sensors to deploy, at what frequency to sample data,  what information they provide at what different times of day, and seasons, with what different levels of occupant activity to expect, and with which soft sensors\cite{agarwal2016observability} to combine and make inferences with,  simulations will take an increasingly important role in controlling and testing the scope of inferences in buildings. \edits{Methods like this demonstrate that simulations have the potential to serve as a stand-in for domain experts. If experts can digitize the knowledge of specific sensor placements for building commissioning, for example, they can enable accessible and code-compliant occupant privacy protection designs while further providing availability to interact with other simulations using the same building model.}

\paragraph{Broader Impacts}
The work we have completed represents both predictive model tasks, where we anticipate the use-case of the occupant before installing the sensors, but also a step towards reducing the gap between the digital twin and the original twin. Sensor installations can take advantage of more than just their placement concerning common building elements such as floors, walls, and doors. Sensor placements can also benefit from being aware of other sensors in the context. The workflow we demonstrate allows for explorations in designing buildings that can be more effectively commissioned with fewer sensors. Developing metrics to quantify possible inferences also provides an additional avenue for designers and researchers to consider user privacy. For instance, there could be dedicated ``silent zones" where sensors cannot detect any occupant activity as protected by the laws of physics. Simulations can be an effective tool to compete with the scale of sensor developments because they can protect their users from scenarios that have yet to happen and inform and adjust models using real data. While the digital and original twins divide is shortening, we consider their distinct identities to carry certain benefits. For example, digital twins can be operated ``offline" to explore reactive and predictive scenarios that inform on the optimal corrective action without interfering with the operations in the real system. However, this is not to say that simulations will not also be an increasingly important part of the operations in real-time systems. \edits{Decisions to navigate the potential split incentives between the building operator and the occupant will likely depend on circumstance and might require routine updates to support new management and new tenants. A building operator might install light sensors in all rooms to avoid the complexity of inference but expose information about the occupant's kitchen, living room, and bathroom use they consider private. Similarly, an occupant might install one sensor to understand lighting in the living room but accidentally leak the lighting states of other rooms to the building operator. Decisions regarding which data should be hidden for privacy or which data should be available for utility would require coalescing of ideologies regarding ownership of space, ownership of data, ethics, among other considerations. Regardless of the perspective, the first step is showing in a data-driven and reproducible manner where a sensor installation theoretically lies on the privacy-utility spectrum.}

\paragraph{Limitations}
\edits{One limitation of this study was that it was conducted on an older residential unit, where buildings might not reflect a more modern understanding of the efficient usage of light fixtures. The privacy implication of sensor use in public and semi-public situations such as offices and libraries could have a broader impact on the number of people affected. Another limitation is our assumption of snapshot views in simulations. We did not incorporate time (analysis of signals instead of values) such as through bulb response functions \cite{sheinin2017computational}. This can limit the ability for simulations to reflect the time in-between snapshots and the additional inferences that can be drawn as a result of more realistic sample rates}. Another limitation is that we only collected light intensity levels but did not look further into other properties of light, such as lighting colors. We suspect colors can be an important avenue to disambiguate lighting signals further. For instance, the individual lights might be able to be first filtered by color, reducing the total number of possible combinations and collisions that could happen. Another limitation is that the involvement of some furniture has the potential to alter the indoor environment. For example, having a mirror on the wall can drastically alter the luminous environment, similar to having light-absorptive materials on the floors. Finally, our work does not address the difficulty of constructing a representative building model, nor the potential diminishing returns of modeling the environment in more realistic detail. While numerous benefits can be achieved with an informative building model, the cost of building a representative model can eventually outweigh the demand to protect an occupant's privacy. \edits{The cost of the building model can be further exacerbated when more computation time is required to calculate physical interactions in the space, such as increases in the number of bounces for lighting simulations or number of particles in Computational Fluid Dynamics (CFD) simulations.}


\section*{Conclusion} \label{sec:conc}
We demonstrate a theoretical framework for indoor activity inference selection through simulation experiments and real-world sensor placements. We show how simulations can quantify inferable occupant activities \edits{using a distinctness score} and how to find a mathematically minimum set of sensor positions required to detect them \edits{by applying the concept of set cover}. The resulting metrics to quantify distinguishable activities enable future sensor deployments to consider building geometry better and limit potential sensor data overreach. We anticipate using sensor positioning paired with building simulations to grow as an essential technique for researchers to navigate the privacy-utility trade-off for the smart buildings of tomorrow. 

\section*{Data availability Statement}
The datasets generated during and/or analyzed during the current study are available from the corresponding author upon reasonable request.

\bibliography{sample}



\section*{Acknowledgements}
This work was partly supported by the Virginia Commonwealth Cyber Initiative (CCI) grant, and the National Science Foundation (NSF) Grant No. 1823325.

\section*{Author contributions statement}
A.W., B.C., A.H. contributed to ideation. A.W. conducted the experiments and analyzed the results. All authors reviewed the manuscript. 

\section*{Additional information}
The author(s) declare no competing interests.




\end{document}